\documentclass[aps,prb,twocolumn,showpacs,amsmath,amssymb]{revtex4}
\usepackage{epsfig}
\usepackage{dcolumn}
\usepackage{bm}
\usepackage{amsmath}
\usepackage{amsfonts}
\usepackage{amssymb}
\usepackage{graphicx}%
\newcommand{\beq}{\begin{eqnarray}}
\newcommand{\eeq}{\end{eqnarray}}

\newcommand{\omi}{(\omega)}

\begin{document}
\title{DC four point resistance of a double barrier quantum pump}
\author{Federico Foieri$^1$, Liliana Arrachea$^{1,2}$ and Mar\'{\i}a Jos\'e  S\'anchez$^3$.}
\affiliation{$^1$Departamento de F{\'i}sica ``J. J. Giambiagi" FCEyN, Universidad de Buenos 
Aires, 
Ciudad Universitaria Pab.I, (1428) Buenos Aires, Argentina,\\
$^2$ BIFI, Universidad de Zaragoza, Pedro Cerbuna 12, 50009, Zaragoza, Spain,\\
$^3$Centro At\'omico Bariloche and Instituto Balseiro, Bustillo 9500 (8400), Bariloche, 
Argentina.}
\date\today 
\begin{abstract}

We investigate the behavior of the dc voltage drop in a periodically driven 
double barrier structure (DBS) sensed by voltages probes that 
are weakly coupled to the system. 
We find that  the four terminal resistance $R_{4t}$ measured with  the probes located 
outside the DBS
results identical to the  resistance  measured in the same structure  under a 
stationary bias  voltage difference  between left and right reservoirs. 
This result, valid beyond the adiabatic pumping regime, can be taken as an indication
of the universal character of $R_{4t}$ as a measure of the resistive properties of 
a sample, irrespectively of the mechanism used to induce the transport.
\end{abstract}
\maketitle
Quantum transport induced by time dependent fields attracts presently an impressive amount of  
research. 
A phase coherent conductor subjected to two periodically varying voltages becomes a paradigmatic 
example of a quantum pump, in which a dc current can be generated in the absence of a net external 
bias \cite{SMCG99,pump,mobu,lilip}.

After the works of Landauer and Buttiker,\cite{L70,BU8688} the four point resistance $R_{4t}$ is considered 
as the proper
measure of the genuine resistive behavior of a mesoscopic sample, free from the effects of  the contact 
resistance.
 Several theoretical works have been devoted to 
study the  details of the voltage drop between the contacts in systems where 
 the  transport is induced  by means of a stationary dc voltage bias \cite{DP90}. Furthermore, the striking feature
 that $R_{4t}$ can be negative in a coherent conductor has been experimentally observed  in semiconductors \cite{pic} and 
 in carbon nanotubes  \cite{gao}.
 However, the behaviour of  $R_{4t}$ in the case of a quantum pump  has not been so far analyzed. 

\begin{figure}
\includegraphics[width=0.9\columnwidth,clip]{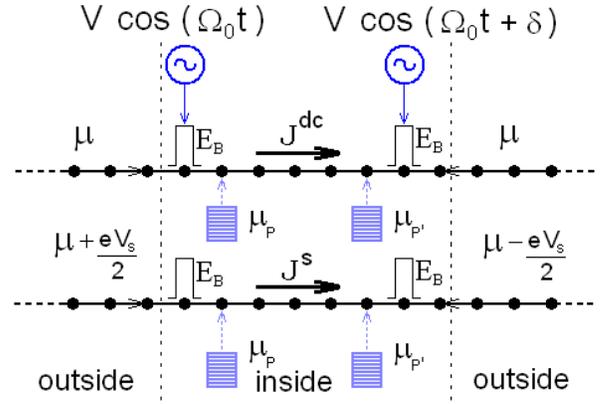}
\caption{\label{fig1} (Color on-line)
Scheme of the set up. The central device is a wire with two  barriers of height $E_B$ connected to $L$ and $R$ reservoirs. Two  voltage probes 
$P$ and $P^{'}$ sense the voltage drop. 
Upper plot: Pumping set up in which a dc current $J^{dc}$ is induced by two local   ac voltages. The
$L$ and $R$ reservoirs are at  the same chemical potential $\mu$. 
Lower plot: Stationary set up, in which the current $J^{s}$ is induced by a dc voltage difference $V_s$ between $L$ and $R$ reservoirs. See text for more details.}
 \end{figure}

The aim of  the present work is to investigate to what an extent the concept of $R_{4t}$ could depend on the 
underlying driving mechanism.
 To this end, we consider as a model of the quantum pump, a quantum wire coupled to left and right reservoirs 
 at a fixed chemical potential and   with two  narrow gates 
 to which oscillating voltages are applied with a phase-lag. This set up (see Fig.\ref{fig1} upper plot) mimics the actual 
 double barrier structure (DBS) used in Ref.\onlinecite{SMCG99} where two of such ac potentials 
 were applied at the walls confining a quantum dot. Experimentally, the dc response  is actually 
 inferred from the measurement of the voltage drop between two extra probes,  one located at the left and 
 the other at the right of the DBS.
Accordingly, we consider  non-invasive voltage probes  weakly  coupled to the wire. The chemical potential 
$\mu_i$ ($ i= P,P^{'}$) of  each probe is adjusted to maintain  zero net  current through the respective contact. Our goal
is to investigate the dc four terminal resistance for the quantum pump defined
as the dc voltage drop $\Delta \mu_{PP'}$ sensed by the two probes $P, P'$
connected at two arbitrary points along the sample, divided by the 
dc component of the current $J^{dc}$ flowing through the device :
\begin{equation}
R_{4t}= \frac{\Delta \mu_{PP'}}{J^{dc}}.
\end{equation}  
We  will also compare this quantity  with the four terminal resistance
obtained when the current through the DBS is induced by a slight stationary voltage difference $V_s$ between the left ($L$) and right ($R$) reservoirs, as indicated in the lower panel of
Fig.\ref{fig1}.   
 
For  sake of clarity we start considering just one
voltage probe, which is modeled as a third reservoir coupled to the central system at  
the position $P$ (the extension to more
probes is trivial). The corresponding Hamiltonian for the full system reads: 
\begin{eqnarray}
H  & = &  H_{leads} + H_P  + H_C(t) -w_L (a^\dagger_L c_1 + H.c.) \nonumber \\
& & -w_R (a^\dagger_R c_N + H.c. ) -w_P (a^\dagger_P c_p + H.c.) 
\label{ham1d}
\end{eqnarray}
with $H_C(t)$  denoting the Hamiltonian for the central piece that we model as a 1d tight-binding chain of length $N$ with two barriers located
at sites $A$ and $B$:
\begin{eqnarray}
& & H_C (t)  =  
V \cos(\Omega_0 t + \delta ) c_{A}^\dagger c_{A} 
 +\sum_{l=1}^N \varepsilon_l c^\dagger_l c_l \nonumber \\
& &  - w_h \sum_{l=1}^N (c_l^\dagger c_{l+1} +H.c) +
V \cos(\Omega_0 t ) c_{B}^\dagger c_{B},
\label{ham2}
\end{eqnarray}
with $w_h$ the hopping parameter and the profile $\varepsilon_A=\varepsilon_B=E_b$, $\varepsilon_l=0,
l=1,\ldots,N \neq A, B$, defining a double barrier structure.
The time dependent  ac potentials act locally at the position of the barriers and have amplitude $V$, 
frequency $\Omega_0$  and oscillate with a phase difference $\delta$. 
We denote with $H_{leads}$ the Hamiltonians of 
two semi-infinite tight-binding chains with hopping $w_l$ at the same chemical potential $\mu$, 
which play the role of the L and R 
 reservoirs. These two leads are
connected to the central device at sites $1, N$ respectively.
 Similarly, $H_P$ is the hamiltonian of the voltage probe $P$ that we also model as a particle 
reservoir with a
chemical potential $\mu_P$ that is fixed to satisfy the condition of net zero dc current through its
contact to the central system \cite{BU8688}.
The contacts between the central system and the $L$ and $R$ leads and the probe $P$ are described by 
the last three terms of Eq.(\ref{ham1d}), where the fermionic operators 
$a_\alpha$ ($\alpha= L, R, P$) denote degrees of freedom for the $L, R$ and $P$ reservoirs 
respectively.

We employ the formalism of Keldysh non equilibrium Green's functions, 
which is  a convenient  tool
in transport theory on  multiterminal structures driven by time-periodic fields.
Following Ref.\onlinecite{lilip} we use the Floquet representation
 $G^R_{l,l'}(t,\omega)=
\sum_{k=-\infty}^{\infty} e^{- i k \Omega_0 t} {\cal G}_{l,l'}(k,\omega)$,
where $G^R_{l,l'}(t,\omega)$ is the Fourier transform  with respect to $t-t'$ of the retarded Green's
function.
The dc component of the  charge current flowing through the contact between the central system
and the probe  $P$,  can be written (in units of $e/h$) as \cite{lilip}:
\begin{eqnarray}
\!\!& & J_P^{dc}=
\sum_{\alpha=L,P,R}
\sum_{k=-\infty}^{\infty} \int_{-\infty}^{\infty} 
\frac{d\omega}{2\pi} \lbrace \Gamma_{\alpha}
\omi \Gamma_P(\omega+k\Omega_0) \nonumber \\
\!\!& & |\mathcal G_{l_{P},l_{\alpha}}(k,\omega)|^2 
\left[f_{\alpha}\omi - f_{P}(\omega+k\Omega_0 ) \right] \rbrace,
\label{jdcp}
\end{eqnarray}
where $l_{\alpha}$ are the sites of the central system at which the reservoirs
$\alpha= L, R, P$ are attached, while $\Gamma_{\alpha}(\omega) = |w_{\alpha}|^2 \rho_{\alpha}(\omega)$ is the spectral function
 associated to the self-energies due to the coupling to these reservoirs,
$\rho_{\alpha}(\omega)$ is the corresponding density of states and $f_{\alpha}(\omega)=1/(e^{\beta_{\alpha}(\omega - \mu_{\alpha})}+1)$ is the Fermi function of  the reservoir $\alpha$, which we assume to
be at the temperature $1/\beta_{\alpha}=0$. 

The voltage profile sensed by the probe can be exactly evaluated under general conditions
from  the solution $\mu_P$ that satisfies $J_P^{dc}=0$ in the above expression. It is, however, instructive to analyze first the case
of  low driving frequency $\Omega_0$ and small pumping amplitude $V$, which corresponds to the so called adiabatic pumping regime \cite{pump}.
As we already mentioned,  we  are considering ``non-invasive probes''. This corresponds to probes weakly 
coupled to the central system, in such a way that they do not introduce neither inelastic nor elastic scattering processes for the electronic propagation 
between $L$ and $R$ reservoirs. Below we  derive an analytical expression,  valid under these conditions, for the  voltage profile $\mu_P$. 

\begin{figure}
\includegraphics[width=0.9\columnwidth,clip]{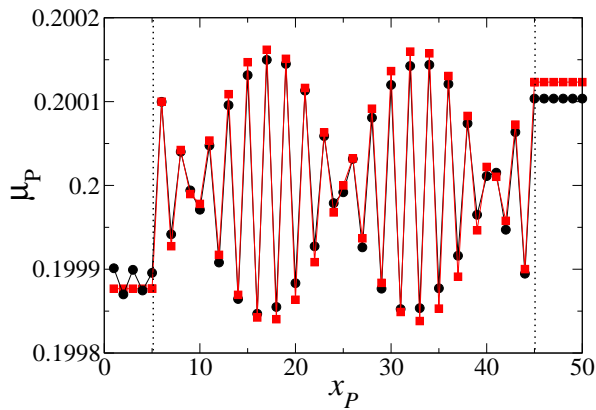}
\caption{\label{fig2} (Color on-line)
Local voltage $\mu_P$ sensed by the voltage probe $P$ as a function of the probe position $x_P$ along a DBS composed 
by $N=50$ sites with two  barriers of height $E_B= 0.2$ located at  $x_A=5$ and $x_B=45$ as indicated by the vertical dashed lines. 
The pumping parameters are $V=0.01$,  $\Omega_0=0.01$ and $\delta=\pi/2$.
 Red squares correspond to Eq.(\ref{mupad}), the analytical solution for the adiabatic pumping regime and  
 a weakly connected probe.
Black circles correspond to the exact numerical solution obtained equating Eq. (\ref{jdcp}) to zero with
 $w_P=0.01$. The chemical potential is $\mu=0.2$, which corresponds to $k_F =1.47$.}
\end{figure}

For low $V$, the perturbative solution of the Dyson's equation up to the second order in this parameter 
 contains only the following Floquet components that contribute to the dc current \cite{lilip}:
\begin{eqnarray}\label{gfloq}
&& \mathcal G_{l,l'} (\pm 1,\omega)  \sim \frac{V}{2} [ G^0_{l,A}(\omega \mp \Omega_0 )G^0_{A,l'}(\omega) \nonumber \\
&& + e^{\pm i\delta} G^0_{l,B}(\omega \mp \Omega_0) G^0_{B,l'}(\omega)  ] .
\end{eqnarray}
For weak coupling to the probes, Eq.(\ref{jdcp}) is evaluated with 
Green's functions  up to the 1st order in $w_P$. This corresponds to consider  the functions $ G_{l,l'}^0(\omega)$ in
(\ref{gfloq}), as the equilibrium retarded Green's functions 
of the central system attached only to the   $L$ and $R$ reservoirs.
 For perfect matching to the reservoirs 
($w_L=w_R=w_l=w_h$) and for barriers with low amplitude $E_B \leq w_h$, these functions can be written in the following simple form:  
$G^0_{l,l'}(\omega)=g_{l,l'}(\theta) + E_B \sum_{j=A,B}g_{l,j}(\theta)g_{j,l'}(\theta)$, with 
$g_{l,l'}(\theta)= i e^{-i|l-l'| \theta}/(2 w_h \sin \theta)$, being $\omega= 2 w_h \cos \theta$ \cite{SO89}.
Using them to evaluate (\ref{gfloq}), substituting the result in (\ref{jdcp}) and considering the
adiabatic ($\propto  \Omega_0$) contribution in the resulting $J^{dc}_P$ we get two different results
depending on the place at which the probe is connected:
\begin{eqnarray} \label{mupad}
 \mu_P && = \mu \pm \Omega_0  V^2 \sin{ \delta} [\alpha^o(k_F)  \nonumber \\
&& + E_B \beta^o(k_F)], \;\;\;\;\;x_P > x_B, x_P <x_A,\nonumber \\
 \mu_P && = \mu +\Omega_0 V^2 \sin{ \delta} [ \alpha^i(k_F, x_P) \nonumber \\
&&
+ E_B \beta^i(k_F,x_P)], \;\;\;\;\; x_A<x_P <x_B,
\end{eqnarray}
with $x_j (j=A, B, P)$ denoting the position of the barriers and probe  in units of the lattice parameter of the tight-binding model.
The upper and lower  sign of the first identity corresponds, respectively, to the voltage probe located at  the left  ($x_P <x_A$)  and right 
($ x_P > x_B$) side  of the  DBS,
while the second identity corresponds to the voltage probe located between the two barriers.
We have defined the Fermi vector (in units of the lattice parameter) as $k_F \equiv \theta(\mu)$ as well as the following functions: 
\begin{eqnarray}\label{albetgam}
\alpha^o (k_F)&&= \dfrac{ \sin[2 k_F(x_A-x_B)]}{2 (w_h \sin k_F)^2},   \nonumber \\  
 \beta^o (k_F)&&= \dfrac{\sin^{2} [k_F(x_A-x_B)]}{4 ( w_h \sin k_F)^3},   \nonumber \\
\alpha^i (k_F,x_P)&&= \sin[k_F(2x_P - x_A-x_B)]\alpha^o(k_F),   \nonumber \\
\beta^i(k_F,x_P)&&=\sin[k_F(2x_P -x_A-x_B)]  \beta^o (k_F),    
\end{eqnarray}
where the superscripts $o, (i)$ stress that the probe senses points outside (inside) the region where all the scattering processes 
(dynamical as well as stationary) take place.
In the simple model we are considering, with a perfect matching between the central system and the reservoirs, this coincides with the DBS,
as indicated in Fig. 1.
Equations (\ref{mupad}) and (\ref{albetgam}) tell us that the local voltage sensed by a probe is constant outside the region defined by the DBS, while 
it presents the characteristic pattern of  Friedel oscillations \cite{DP90}
with a period $2 k_F$  at  positions lying between the two oscillating barriers.
Fig. \ref{fig2} shows the benchmark of the analytical result, Eq.(\ref{mupad}), against the exact voltage profile obtained numerically from  Eq. 
(\ref{jdc}) in the regime of weak $V$, $\Omega_0$ and $w_P$, and a moderate $E_B$. A good agreement of the qualitative behavior is observed. 
In particular, the exact profile $\mu_P$ exhibits  Friedel oscillations
with period $2 k_F$ as a function of the probe position $x_P$ as predicted by Eq.(\ref{mupad}) and only a slight disagreement is found in the 
amplitude of the envelope function.  

To  the lowest order of perturbation in the coupling constant $w_P$, the  effect of an additional second 
voltage probe $P'$ is completely uncorrelated from the first one, since the associated  interference effects involve second order processes in  $w_P$. At this level of approximation let us call $\mu_{P'}$ the  local voltage sensed by the additional probe at $P'$
and $\Delta \mu_{PP'}\equiv \mu_{P'} - \mu_P$ the corresponding voltage drop. 
In a set up  with the   probe P (P') located at the left (right) side of the DBS, the voltage drop between both probes is from Eq.(\ref{mupad}):
\begin{eqnarray}\label{dmuo}
&&\Delta^o \mu_{PP'} = 2 \Omega_0 V^2 \sin\delta [\alpha^o(k_F) + E_B \beta^o(k_F)].
\end{eqnarray}
Another possible measurement corresponds to locate the voltage probes P and P' inside the DBS. In this case,
 the voltage drop between the two probes  explicitly depends on the probe positions $x_P$ and $x_{P'}$ as follows:
 \begin{eqnarray}\label{dmui}
\!\!\!\!\!\!\!&& \Delta^{i} \mu_{PP'} = 2 \Omega_0  V^2 \sin \delta
\left[\dfrac{\alpha^o(k_F)+E_B \beta^o(k_F)}{\sin(k_F(x_A-x_B))} \right] \times  \\
\!\!\!\!\!\!\!&& \cos\{k_F[(x_P-x_A)+(x_{P'}-x_B)]\} \sin[k_F(x_{P'}-x_{P})]\;,\nonumber
\end{eqnarray}
where, as before, we have employed the superscripts $o, (i)$  to distinguish configurations with the probes outside (inside) the DBS.

Under the conditions assumed in the derivation of Eq.(\ref{mupad}), i.e. low $V, \Omega_0, w_P$ and $E_B$, the dc current flowing through the DBS reads:
\begin{equation} \label{jdc}
J^{dc} \cong 2 \Gamma_L^0 \Gamma_R^0 \Omega_0 V^2 \sin\delta 
\left[ \frac{\alpha^o(k_F) + E_B \beta^o(k_F)}{(2 w_h \sin k_F)^2} \right],
\end{equation}
 with $\Gamma_\alpha^0 \equiv  \Gamma_\alpha^0(\mu), (\alpha=L,R)$. 
We can  now compute the dc four terminal resistance $R_{4t}$ in  an adiabatic weakly driven pumping process.
For a set up in which the probes are located outside the DBS ($ x_P< x_A,x_B < x_{P'}$) it reads:
\begin{equation} \label{r4o}
R^{o}_{4t}= \frac{\Delta^{o} \mu_{PP'}}{ J^{dc}} = \dfrac{(2 w_h \sin k_F)^2}{\Gamma_L^0 \Gamma_R^0}\;,
\end{equation} 
while it is:
\begin{eqnarray} \label{r4i}
R^{i}_{4t} & = & 
 \!\!\frac{\Delta^{i}\mu_{PP'}}{ J^{dc}} \nonumber \\
& = & E_B R^{o}_{4t} \; \!\!\frac{\sin[k_F(x_{P'}-x_P)]}{\sin[k_F(x_A-x_B)]}\nonumber \\
& & \times \cos[k_F(x_P-x_A+ x_{P'}-x_B)] \; ,
\end{eqnarray}
for a set up in which the probes are located inside the DBS ($ x_A < x_P < x_{P'}<x_B$).

We now turn to  compare the value of  $R_{4t}$ obtained for   the quantum  pump with the resistance of
the same DBS when  the transport  is induced through a stationary bias voltage $V_s$ established by a 
difference in the  electrochemical potentials of $L$ and $R$ reservoirs $\mu_L = \mu + e V_{s}/2$ 
and $\mu_R = \mu - e {V_s}/2$, as is depicted in the lower panel of Fig.\ref{fig1}. 
Under the  conditions of   non-invasive probes  and for linear response in $V_{s}$, it
is possible to implement the same kind of perturbative procedure as before, but for the stationary 
Green's functions. Assuming again perfect matching between
the DBS and the $L$ and $R$ reservoirs and  $E_B < w_h $, we derive the following simple expression for 
the current
flowing through the device,
\begin{equation}\label{js}
J^{s} = \frac{\Gamma^0_L \Gamma^0_R V_s}{(2 w_h \sin k_F)^2} \;, 
\end{equation}
which is the stationary counterpart of Eq.(\ref{jdc}).
>From the above expression we compute  ${V_s}/ {J^s}$  and  arrive  immediately  to  the important 
identity:
\begin{equation}\label{id1}
R^{o}_{4t} \equiv \frac{V_s}{J^s} \; ,
\end{equation}
which tells  that the dc four point resistance measured in the quantum pump when the probes are 
connected outside the DBS  $R^{o}_{4t}$ (Eq.(\ref{r4o})) exactly coincides with  the  {\em total resistance } of the structure 
measured  under stationary bias, provided that the driving condition corresponds to linear response in the stationary setup and the adiabatic 
regime in the pumping setup. At this point, it is important to recall that in the outside configuration the two probes enclose 
the whole region where all the scattering processes and, therefore, the full voltage drop $V_s$ applied in the stationary setup takes place.

On the other hand, the counterpart of Eq. (\ref{r4i}) for the stationary configuration reads:
\begin{equation}\label{id2}
R^{s,i}_{4t} \equiv R^{i}_{4t} \frac{\sin [k_F(x_A-x_B)] }{\sin k_F},
\end{equation}
which means that inside the DBS the dc resistance for the pumping setup differs from that under stationary driving just in a geometrical factor.

\begin{figure}
\includegraphics[width=1.0\columnwidth,clip]{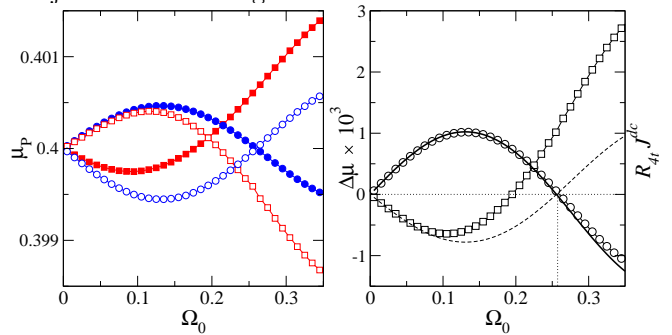}
\caption{\label{fig3} (Color on-line)
Left panel: Exact local voltage $\mu_P$ sensed at $x_P= 2, 14$ (empty symbols) and
$x_{P'}=26, 38$ (solid symbols),
as a function of the pumping frequency $\Omega_0$ obtained  numerically equating Eq. (\ref{jdcp}) to zero. Circles (squares) correspond
to points outside (inside) the DBS. Parameters are: $\mu=0.4$, $N=40$, $E_B= 0.2$, $x_A=10$ and $x_B=30$. 
Right panel: The corresponding potential drops  $\Delta^{o,i} \mu_{P P'}$.
The products $R^{o,i}_{4t} J^{dc}$ (solid and dashed lines, respectively) between the exact current $J^{dc}$ and the
resistances evaluated from Eqs. (\ref{r4o}) and (\ref{r4i}), respectively. The vertical dotted line indicates the frequency of the pumping 
at which current inversion occurs.}
\end{figure}

Besides the relevance of the  analytical results it is valuable  to analyze the response of the system beyond the adiabatic pumping condition. For that purpose we have performed extensive numerical calculations of the dc current flowing through the system $J^{dc}$ and  the potential drop sensed by the voltage probes along the DBS, as a function of the pumping frequency $\Omega_0$.
In the left panel of Fig.(\ref{fig3}) we plot the  chemical potential $\mu_P$ as a function of  
$\Omega_0$ for different locations of the probe $x_P$. In addition, in the right panel we have 
plotted the voltage drop $\Delta \mu= \mu_{P'} - \mu_P$ between two probes
located outside the DBS (circles) and inside the DBS (squares). 

When the frequency  $\Omega_0$ is close to the energy difference between two neighbouring levels
of the DBS, the latter become mixed by the pumping potential which causes an inversion in the sign 
of the dc current.
A rough estimate for the frequency at which
such a resonant condition is achieved
in the example of Fig. 3 casts $\Omega_0 \sim 0.22$. In good agreement, we find an
inversion in the sign of $\Delta^o \mu_{PP'}$  for probes connected
outside the DBS at  $\Omega_0 \sim 0.25$ (plots in circles of Fig. 3). 
Moreover, the voltage drop between two points outside the DBS
 in this case results identical to the product of the dc current $J^{dc}$ times the value of the four point resistance $R^{o}_{4t}$ obtained in the adiabatic pumping regime in  Eq.(\ref{r4o}). In other words,
our results indicate that even
 for pumping frequencies
$\Omega_0$ beyond the adiabatic regime, 
 it is still possible to unambiguously define  $R^{o}_{4t}$ as the value obtained under the adiabatic approximation (Eq.(\ref{r4o})). 
On the other hand, the voltage drop measured inside the DBS coincides with the product of $R^{i}_{4t}$ times $J^{dc}$ only within the adiabatic pumping regime, {\em i.e} when $\Delta^{i} \mu_{P P'}$ depends linearly on $\Omega_0$.

In conclusion we have shown that the four terminal resistance $R^{o}_{4t}$ measured in a pumping set up  
{\em for probes located outside the DBS} coincides with the total resistance of the structure measured 
 under stationary bias. This result can be taken as an indication of the universal character of $R^{o}_{4t}$ as a concept to
characterize the resistive properties of a system. Our calculation could be extended to the case where the probes themselves are subjected to 
ac voltages.

We acknowledge support from CONICET, PICT 0313829 (MJS), UBACYT (FF and LA), Argentina.


\begin{thebibliography}{99}
\bibitem{SMCG99}M. Switkes, {\em et al}
   Science {\bf 283}, 1905 (1999).
\bibitem{pump} 
 L. DiCarlo,  {\em et al},
    Phys. Rev. Lett. {\bf 91}, 246804 (2003);
M. D. Blumenthal, {\em et al }, Nature Physics
{\bf 3}, 343 (2007);
    P. W. Brouwer, 
    Phys. Rev. B {\bf 58}, R10135 (1998);
   I.L. Aleiner and A.V. Andreev,
   Phys. Rev. Lett. {\bf 81}, 1286 (1998);
    F. Zhou, {\em et al },
    Phys. Rev. Lett. \textbf{82}, 608  (1999);
    J.E. Avron,  {\em et al },
    Phys. Rev. Lett. {\bf 87}, 236601 (2001);
    V. Kashcheyevs {\em et al},
    Phys. Rev. B {\bf 69}, 195301 (2004); J. Splettstoesser, {\em et al}, Phys. Rev. Lett.  {\bf 95}, 246803 (2005);
S. Kim {\em et al.}, Phys. Rev. B {\bf 73}, 075308 (2006); 
E. Faizabadi, Phys. Rev. B {\bf 76}, 075307 (2007);
   V. Moldoveanu {\em et al },  Phys. Rev. B {\bf 76}, 165308 (2007);  Amit Agarwal {\em et al.}, Phys. Rev. B {\bf 76}, 035308 (2007).
\bibitem{mobu}  M. Moskalets and M. B\"uttiker,  Phys. Rev. B {\bf 66}, 205320 (2002);
 Phys. Rev. B {\bf 69}, 205316 (2004);  Phys. Rev. B {\bf 78}, 035301 (2008).
\bibitem{lilip} L. Arrachea, Phys. Rev. B {\bf 72}, 125349 (2005); 
L. Arrachea and M. Moskalets,  Phys. Rev. B {\bf 74}, 245322 (2006).  
\bibitem{L70} R. Landauer, Philos. Mag. {\bf 21}, 863 (1970).
\bibitem{BU8688}  M. B\"uttiker, Phys. Lett. {\bf 57}, 1761 (1986). M. B\"uttiker, IBM J. Res. Dev, 
{\bf 32}, 317 (1988).
\bibitem{DP90}  J. L. D'Amato and H. M. Pastawski, Phys. Rev. B {\bf 41}, 7411 (1990); V. A. Gopar {\em et al., ibid} 
{\bf 50}, 2502 (1994); T. Gramespacher and M. B\"uttiker, Phys. Rev. B {\bf 56}, 13026 (1997);
L. Arrachea {\em et al}, Phys. Rev. B {\bf 77}, 233105 (2008).
\bibitem{pic}
R. de Picciotto, et al , Nature {\bf 411}, 51 (2001).
\bibitem{gao}
B. Gao, et al , Phys. Rev. Lett. {\bf 95} 196802 (2005).
\bibitem{SO89} F. Sols at al., J. App. Phys. {\bf 66}, 3892 (1989). 


\end{thebibliography}
\end{document}